\documentclass[aps,prl,10pt,twocolumn,notitlepage,floatfix,superscriptaddress,amsmath,amssymb,tightenlines]{revtex4-1}

\usepackage{graphicx}
\usepackage{textcomp} 
\usepackage{graphicx,color}
\usepackage{amssymb}
\usepackage{enumerate}

\newcommand{\NTT}{NTT Basic Research Laboratories, NTT Corporation, 3-1 Morinosato-Wakamiya, Atsugi, Kanagawa, 243-0198, Japan.}

\newcommand{\NICT}{National Institute of Information and Communications
Technology, 4-2-1,
Nukuikitamachi, Koganei-city, Tokyo 184-8795 Japan}

\begin{document}
\title{Observation of collective coupling between an engineered ensemble of macroscopic artificial atoms and a superconducting resonator
}

\author{Kosuke Kakuyanagi}\email{kakuyanagi.kosuke@lab.ntt.co.jp}			\affiliation{\NTT}
\author{Yuichiro Matsuzaki}			\affiliation{\NTT}
\author{Corentin D\'{e}prez}			\affiliation{\NTT}
\author{Hiraku Toida}			\affiliation{\NTT}
\author{Kouichi Semba}  				\affiliation{\NICT}
\author{Hiroshi Yamaguchi}				\affiliation{\NTT}
				\author{William J.  Munro} \affiliation{\NTT}
\author{Shiro Saito} 	\affiliation{\NTT}

\date{\today}

\begin{abstract}
 \textcolor{black}{
The hybridization of distinct quantum systems is now seen as an
 effective way to engineer the properties of \textcolor{black}{an entire} system leading to
 applications in quantum metamaterials, quantum simulation, and quantum
 metrology.
 One well known example \textcolor{black}{is} superconducting circuits coupled to
 ensembles of \textcolor{black}{microscopic natural} atoms.
 In such cases, the properties of the
 \textcolor{black}{individual} atom  are intrinsic, and so are
 unchangeable.
 However, current technology
 allows us to
 fabricate
 large ensembles of \textcolor{black}{macroscopic} artificial atoms such as superconducting flux qubits, where we can
 really tailor \textcolor{black}{and control} the properties of individual
 qubits. 
 Here, we demonstrate coherent
 coupling between a microwave resonator and several thousand
 superconducting
 flux qubits, \textcolor{black}{where we} observe a large dispersive frequency shift in the spectrum
 of 250 MHz \textcolor{black}{induced by} collective behavior.
 \textcolor{black}{These results represent the largest number of coupled superconducting
qubits realized so far.}
 Our approach shows that it is
 now possible to engineer the properties of the ensemble,
 opening up the \textcolor{black}{way}
 for the controlled exploration of the quantum many-body system.}
\end{abstract}
\maketitle



\textcolor{black}
{Quantum science and technology have reached a very interesting stage in
\textcolor{black}{their} development
 where we are now beginning to engineer the properties that we require of our
 quantum systems \cite{dowling2003quantum,ladd2010quantum}.
 Hybridization is a core technique in achieving this. \textcolor{black}{An} additional
 (or ancilla) system can be used to \textcolor{black}{greatly} change not only
 the properties of the overall system, but also its environment
 \cite{xiang2013hybrid,benjamin2009prospects,paz2001quantum}.}

 \textcolor{black}{
   \textcolor{black}{Specifically}, a hybrid system composed of many qubits and a common
  field such as cavity quantum electrodynamics
  \cite{blais2007quantum,tsomokos2008fully}
  may provide an excellent
 way of \textcolor{black}{realizing} such quantum engineering, leading to an interesting
 investigation of many-body phenomena including quantum simulations
 \cite{schmidt2013circuit,georgescu2014quantum},
 superradiant phase transitions
 \cite{hepp1973superradiant,wang1973phase,emary2003chaos,lambert2009quantum,lambert2004entanglement,zou2014implementation},
 spin squeezing
 \cite{kitagawa1993squeezed,bennett2013phonon,tanaka2015proposed}, and 
 quantum metamaterials  \cite{zheludev2010road,soukoulis2011past,zheludev2012metamaterials,ricci2005superconducting,lazarides2007rf,anlage2011physics,jung2013low}.
 In this regard, one of the ways to realize such a system is
 \textcolor{black}{to employ} superconducting circuits coupled to
 electron spin ensembles where basic quantum control such as memory operations have been
 demonstrated \cite{zhu2011coherent,saito2013towards,kubo2011hybrid,zhudark2014,matsuzaki2015improvingpra}.
 \textcolor{black}{If we are to} investigate quantum many-body phenomena, we \textcolor{black}{will} need control over the ensemble. In most
 typical superconducting circuit-ensemble hybrid experiments,
 the ensemble has been formed from a collection of either atoms or
 molecules
 with examples including nitrogen vacancy centers
 \cite{zhu2011coherent,saito2013towards,kubo2011hybrid,putz2014protecting},
 \textcolor{black}{ferromagnetic magnons \cite{tabuchi2015coherent},}
 and bismuth
 donor spins in silicon \cite{bienfait2015reaching}.
 In these cases, the
 properties of the atomic ensemble system are basically defined as
 the ensemble is formed, and \textcolor{black}{are} difficult to
 change. \textcolor{black}{However}, our ensembles
 could be composed of artificial atoms such as superconducting
 qubits.}

 \textcolor{black}{
Superconducting qubits are \textcolor{black}{macroscopic} two-level systems
with a significant degree of
design freedom \cite{ClarkeWilhelm01a,makhlin2001quantum}.
Josephson junctions provide the superconducting circuit with
 non-linearity, and we can tailor the qubit properties by changing the
 design of the circuit. Moreover, in contrast to natural atoms that
 are the size of angstrom,  the size of the superconducting circuit is
 around $1-10$ $\mu $m and so
 we can change the properties of individual qubit
 \textcolor{black}{with a time scale of nano-seconds}
 by using
 a control line coupled to each qubit.
Actually, frequency tunability
 \cite{paauw2009tuning,zhu2010coherent}, 
 coherence time \textcolor{black}{control} \cite{reed2010fast}, tunable coupling strength
 \cite{niskanen2007quantum}, engineering selection rules of
 qubit transitions \cite{harrabi2009engineered}, and control of the
 level structure of the qubit \cite{inomata2014microwave},
 have been demonstrated with the
 superconducting qubits. 
 These show the feasibility to engineer the
 properties of the superconducting circuit.}
\begin{figure*}[htb!] 
\begin{center}
\includegraphics[width=1.00 \linewidth]{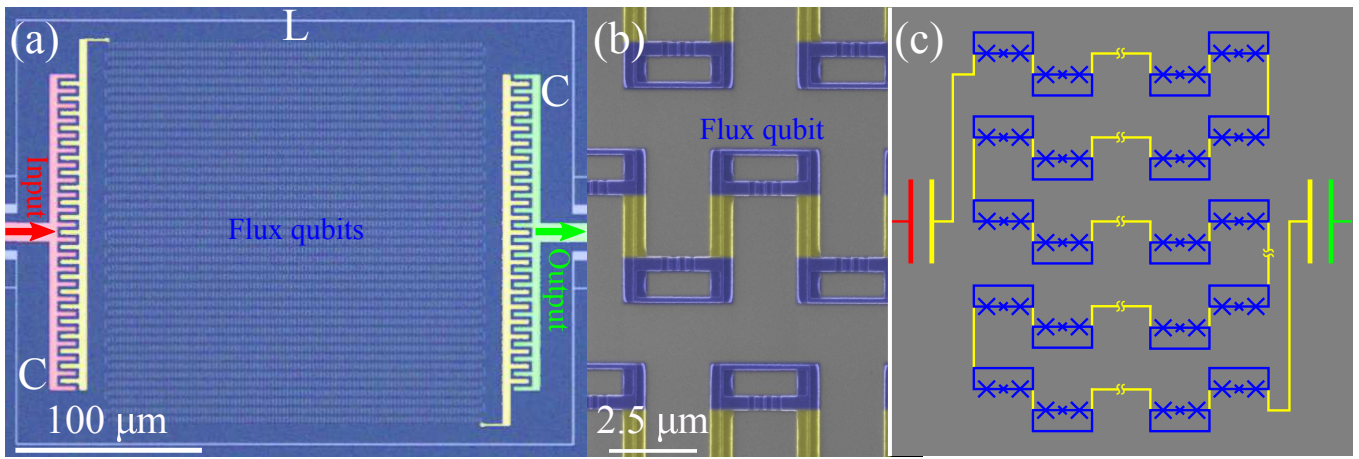} 
\caption{
 Our device is composed of \textcolor{black}{4300} superconducting flux qubits embedded in an LC resonator.
 \textcolor{black}{These} flux qubits are coupled with the resonator via mutual
 inductance. We show \textcolor{black}{in}
(a) an optical microscope image, (b) a scanning electron micrograph (in false colors), and  (c) a schematic view of our device.
 \textcolor{black}{Spectroscopy is performed} on this device by measuring
 \textcolor{black}{the photons transmitted from} the resonator.
 }
 \label{schematic}
\end{center}
\end{figure*}

\textcolor{black}{Besides the tunability, another key issue \textcolor{black}{in terms of observing} interesting quantum phenomena
is how to scale the number of the qubits.
 Collective coupling between three superconducting
transmon qubits and a cavity field has been demonstrated
\cite{fink2009dressed}, and a multi partite entanglement has been
generated with this system \cite{mlynek2012demonstrating}.
Quantum critical behavior has been experimentally investigated with a system where
four superconducting phase qubits are coupled with a resonator \cite{feng2015exploring}.
Also, there \textcolor{black}{has been} an experimental demonstration in
which
\textcolor{black}{20 superconducting qubits are fabricated and 8 qubits
show a collective coupling} \cite{macha2014implementation}.
 However, the number of coherently
 coupled superconducting qubits in the \textcolor{black}{previously reported} demonstration may not be
 large enough for practical
 applications, and so we need to
 extend the system to more qubits.
}

Here, we perform an experiment
where thousands of superconducting flux qubits are coupled with a
superconducting resonator.
Since the resonant
frequency of the superconducting flux qubits \textcolor{black}{is} sensitive \textcolor{black}{to} small
changes \textcolor{black}{in} the fabrication conditions \cite{orlando1999superconducting}, the superconducting
flux qubits suffer from inhomogeneous broadening.
However, \textcolor{black}{the} collective coupling of superconducting flux qubits with a
common resonator can overcome this inhomogeneity, because the coupling strength is enhanced by $\sqrt{N}$ times
due to the collective effect where $N$ is the number of flux qubits \cite{imamouglu2009cavity,wesenberg2009quantum}.
Actually, we have observed a large energy shift of
250 MHz in
the spectroscopy of the resonator. \textcolor{black}{Since the designed
coupling strength between a single flux qubit and a resonator is around 16 MHz, such a large dispersive shift} indicates a collective
enhancement of the coupling strength due to the ensemble of the
superconducting flux qubits. We estimate that thousands of
superconducting flux qubits contributes to this collective coupling.
These results represent the largest number of coupled superconducting qubits
realized so far, \textcolor{black}{and this will lead} to various applications in quantum information
processing.


{\bf{Experimental results}}
\newline \textcolor{black}{
We fabricate a microwave resonator and 4300 flux qubits on
a Si wafer.
The flux qubit consists of a loop interrupted by three
 Al-Al$_2$O$_3$-Al Josephson junctions. We designed the area of one
 junction to be $\alpha $ times smaller
than those of the other two junctions.
  The value of $E_J/E_c$ is 75 in our
  design where $E_J$ $(E_c)$ denotes a Josephson (charge) energy.
  The flux qubits
share an edge with the inductor line of the
resonator.
}

Our experimental setup is shown in Fig. \ref{schematic}.
 \textcolor{black}{We measure the microwave transmission properties of the resonator
system by a network analyzer.  The sample was placed in a dilution refrigerator
operating at 20 mK. We can apply magnetic fields perpendicular to
 the flux qubits, and this can change the operating point of the flux qubit.
 Also, we can change the temperature from below
10 mK (base temperature) to 230 mK by a heater (See methods section for the details).}

\begin{figure*}[t] 
\begin{center}
\includegraphics[width=0.99\linewidth]{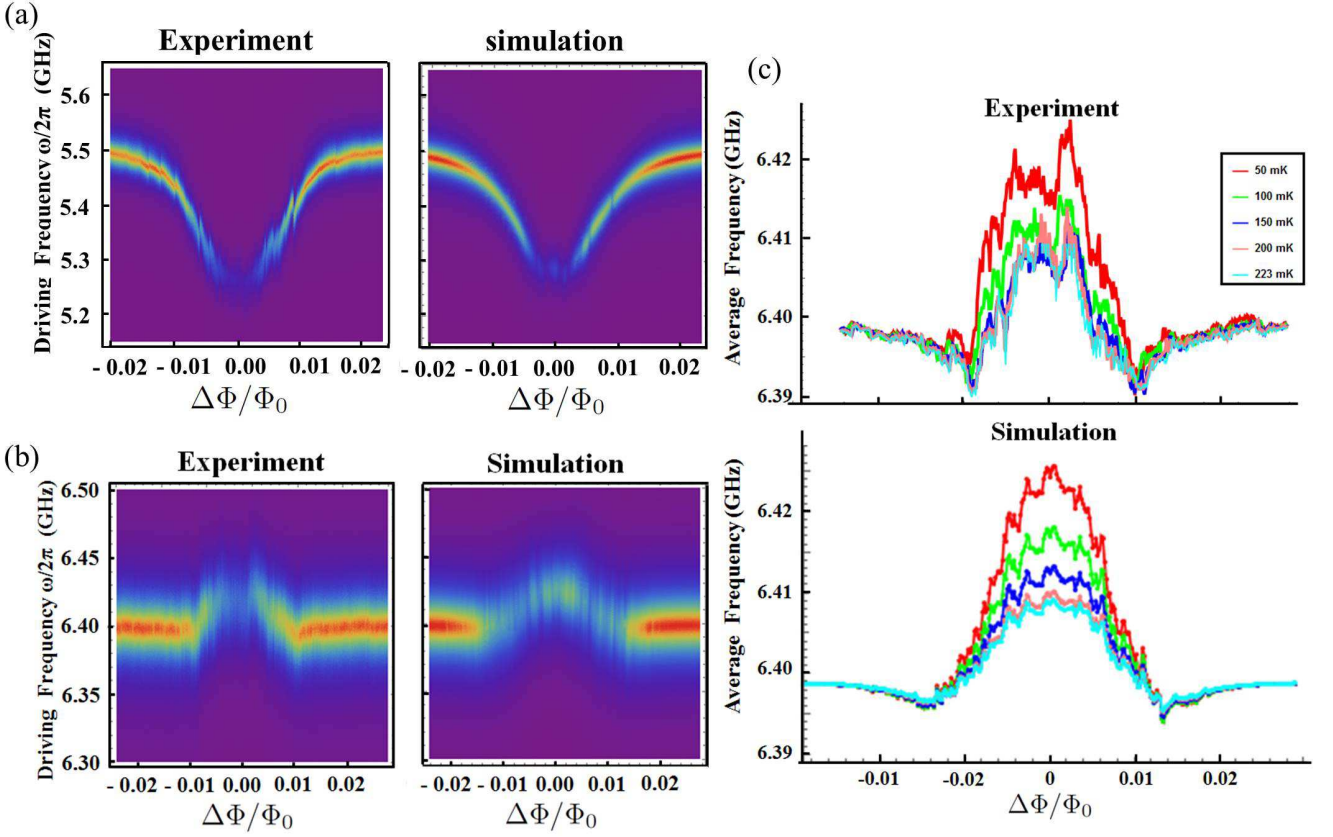}
\caption{Experimental results and numerical simulations of the energy
 spectrum of a microwave resonator coupled to \textcolor{black}{an ensemble} of flux
 qubits. For sample A,
 we used the parameters $N=4300$,  $\bar{\alpha}$= 0.6285, $\bar{\beta }_1$=$\bar{\beta }_2$=1,
 $\sigma _{\text{S}}/\bar{\alpha }=\sigma ^{(1)}_{\text{L}}/\bar{\beta }^{(1)}=\sigma ^{(2)}_{\text{L}}/\bar{\beta }^{(2)}=0.025$, $\omega _{\text{res}}/2\pi =5.5$ GHz,
 $\bar{g}'/2\pi$=14.3 MHz, $\delta \epsilon'_{0}/2\pi =$2.8
 GHz, $\gamma_{\text{qubit}}$/2$\pi$= 50 MHz, and $\gamma_{\text{r}}$/2$\pi$= 13.3
 MHz. For sample B,  we used the parameters $N=4300$, $\bar{\alpha}= 0.7815$, $\bar{\beta }^{(1)}=\bar{\beta }^{(2)}$=1,
 $\sigma _{\text{S}}/\bar{\alpha}=\sigma ^{(1)}_{\text{L}}/\bar{\beta}^{(1)}= \sigma ^{(2)}_{\text{L}}/\bar{\beta}^{(2)}=0.055$, $\omega
 _{\text{res}}/2\pi =6.4$ GHz, $\bar{g}'/2\pi$=9.2 MHz,
 $\delta \epsilon'_{0}/2\pi =$2.6 GHz,
 $\gamma_{\text{qubit}}$/2$\pi$= 50 MHz, and $\gamma_{\text{r}}$/2$\pi$= 12.2
 MHz. \label{sampleb} (c) Temperature dependence of the energy
 spectrum of a microwave resonator coupled to thousands of flux
  qubits. We use the same parameters as those in Fig. \ref{sampleb}.
  From the top,
 we plot the results with $T_{\text{E}}$=50 mK (red), $T_{\text{E}}$=100
 mK (green), $T_{\text{E}}$=150 mK (blue),
 $T_{\text{E}}$=200 mK (pink), $T_{\text{E}}$=223 mK
 (cyan).
 }
 \label{samplea}
\end{center}
\end{figure*}
Spectroscopy \textcolor{black}{was} performed on
the resonator coupled with
thousands of flux qubits for two separate devices (samples A and B)
\textcolor{black}{with different}
designed $\alpha$ values.
By
\textcolor{black}{varying} the driving microwave frequency and applied
magnetic field,
\textcolor{black}{the transmitted photon intensity indicates the resonance frequency of our device.}
In \textcolor{black}{our} experiment, we observed
a \textcolor{black}{resonator} frequency shift
due
to
coupling with the flux qubits.
With sample A, we have observed a large negative resonator-frequency shift of
$\delta \omega_{\text{r}}/2\pi \simeq $250 MHz, \textcolor{black}{and
the width of the spectrum becomes
larger as the frequency shift increases (Fig. \ref{samplea}a.)} 
 \textcolor{black}{On the other hand,}
 for sample B, we observed both a negative and a positive frequency
shift of tens of MHz  (Fig. \ref{samplea}b), \textcolor{black}{and the width of the
spectrum becomes broader as the frequency shifts.}
We also measured the temperature dependence of the resonator
frequency for sample B where we plot the frequency of
the resonator in the spectroscopic measurements (Fig. \ref{samplea}c).
This shows that
an increase in the temperature tends to suppress the frequency shift of
the resonator \textcolor{black}{due to the thermalization of the flux
qubits with small tunneling energies}.
\textcolor{black}{Moreover, on top of the frequency shift, we observe numerous
small energy shifts in the spectroscopy in the
experiments (Fig. \ref{samplea}c). These peaks are reproducible over multiple experiments, and
so they do not correspond to noise. }

{\bf{Theoretical model}}
\newline To understand the mechanism causing the frequency shift of the
resonator,
\textcolor{black}{we model our resonator-qubit ensemble system with a Tavis-Cumming Hamiltonian}
in the rotating
frame of the microwave driving frequency
\cite{tavis1968exact} as 
\begin{eqnarray}
 H&=&H_{S}+H_D+H_I\label{hamiltonian} \\
 H_S&=&\hbar(\omega_{\text{r}}-\omega)\hat{a}^{\dagger}\hat{a}+
  \frac{\hbar}{2} \sum_{j=1}^N
  (\omega _j-\omega)\hat{\sigma}_{z,j} \\
 H_D&=&\hbar \lambda
 (\hat{a}^{\dagger }+\hat{a}) \\
 H_I&=& \hbar \sum_{j=1}^N  g_j(\hat{\sigma}^{+}_{j}\hat{a}+\hat{\sigma}^{-}_{j}\hat{a}^{\dagger})
\end{eqnarray}
where $\hat{a}$ \textcolor{black}{($\hat{a}^{\dagger }$)} represents the annihilation \textcolor{black}{creation} operator of the microwave
resonator, $\lambda $ denotes a microwave driving \textcolor{black}{field
with a frequency of $\hbar \omega $}, $\hbar \omega_{\text{r}}$ denotes the energy of the resonator,
$ \hbar \omega _j =\sqrt{\Delta^{2}_{j}+\epsilon^{2}_{j}} $ denotes the energy
of the $j$ th flux qubit \textcolor{black}{where} $ \Delta_{j}$ is the
tunneling energy of the $j$-th qubit \textcolor{black}{and}
$ \epsilon_{j}=2I_{j} (\Phi^{(j)}_{\text{ex}}-\frac{1}{2}\Phi _0)$
is the energy bias. \textcolor{black}{Here,}  $I_j$ denotes the persistent current of the $j$-th flux qubit, $\Phi _0$
denotes the flux quantum, $\Phi^{(j)}_{\text{ex}}$ denotes the applied
magnetic flux. \textcolor{black}{Next,}
$ g_{j}=\frac{\Delta_{j}}{\sqrt[]{\Delta^{2}_{j}+\epsilon^{2}_{j}}}g'_{j}$
represents the effective coupling strength \textcolor{black}{where} $ g'_{j}$ represents the bare inductive coupling strength
calculated \textcolor{black}{from the} persistent currents and inductance of the devices, and $N$
denotes the number of flux qubits.
We rewrite the energy bias as $ \epsilon_{j}=2I_{j}
(\Phi_{\text{ex}}+\delta \Phi^{(j)}_{\text{ex}} -\frac{1}{2}\Phi _0)
=2I_{j}(\Phi_{\text{ex}}-\frac{1}{2}\Phi _0)+\epsilon_{ \scalebox{0.5}{0}j}$  where  $\Phi_{\text{ex}} $ denotes the average applied
magnetic field and $\epsilon_{ \scalebox{0.5}{0}j}=2I_{j} \delta \Phi^{(j)}_{\text{ex}}$
denotes the energy bias variation caused by the
inhomogeneous magnetic flux $\delta \Phi^{(j)}_{\text{ex}}$.

We can calculate the transmitted photon intensity of the microwave
resonator as follows.
\textcolor{black}{By} solving the Heisenberg equations with
 a weak coupling regime when the system is stationary
 \cite{GZ01b,collett1984squeezing,diniz2011strongly},
 we obtain the transmitted photon intensity $T(\omega )$ as follows
 (See methods section for the details).
\begin{eqnarray}
|T(\omega )|^2&\simeq& \frac{|\lambda |^2}{(\omega-(\omega_{\text{r}}+\delta\omega_{\text{r}}))^2+(\gamma_{\text{r}}+\delta
  \gamma_{\text{r}})^2} \ \ \ \ \ \ \ \ \ 
  \label{theory}\nonumber \\
  \delta\omega_{\text{r}} &=&-\sum_{j=1}^N \frac{g_{j}^2
(\omega_{j} -\omega_{\text{r}}) \cdot  \text{tanh}(\frac{\hbar \omega_{j}}{ 2 k_{B}
T_{\text{E}}})}{(\omega_{j}-\omega_{\text{r}})^{2}+\gamma_{j}^{2}}\\
  \delta\gamma_{r}&=&\sum_{j=1}^N \frac{g_{j}^2
\gamma_{j}}{(\omega_{j}-\omega_{\text{r}})^{2}+\gamma_{j}^{2}}
 \\
   \gamma_{j}&=& (1+(e^{\frac{\hbar \omega _j}{k_BT_{\text{E}}}}-1)^{-1})\gamma
  _{\text{qubit}}
\end{eqnarray}
\textcolor{black}{where $\delta\omega_{\text{r}}$ denotes a frequency
shift of the resonator and $  \delta\gamma_{r}$ denotes the change in
the decay rate of the resonator,}  $T_{\text {E}}$ denotes \textcolor{black}{the} temperature of the environment, $\gamma _{\text{r}}$ denotes \textcolor{black}{the}
decay rate of the resonator,  and $\gamma
_{\text{qubit}}$ ($\gamma _j$) denotes the energy relaxation rate of the flux
qubit at zero (finite) temperature.

The ensemble of flux qubits is affected by inhomogeneous broadening,
  \textcolor{black}{as} it is difficult to make homogeneous junctions, and so
  the area of each junction has a statistical distribution. This results in
  variations in the persistent current and the tunneling energy of
  the flux qubit. 
  \textcolor{black}{The} applied magnetic field is also
  inhomogeneous, which induces the \textcolor{black}{fluctuation} distribution in
  $\epsilon_{\scalebox{0.5}{0}}$.
  We assume a Gaussian distribution for the normalized areas of the smaller
  junction (two larger
  junctions)
  with a mean value of $\bar{\alpha }$
  ($\bar{\beta }_k$ for $k=1,2$) and a standard
  deviation of $\sigma _{\text{S}}$ ($\sigma ^{(k)}_{\text{L}}$ for
  $k=1,2$).
  It is worth mentioning that $g'_j$ and $\epsilon _{\scalebox{0.5}{0}j}$ \textcolor{black}{values}
  are proportional to the $I_j$ \textcolor{black}{value}. So we rewrite these as
  $g'_j=\bar{g} I_j/\bar{I}$ and
  $\epsilon _{\scalebox{0.5}{0}j}=\epsilon'_{\scalebox{0.5}{0}j}I_j/\bar{I}$ where $\bar{I}$ ($\bar{g}$) denotes the average value of the
  persistent current (coupling strength), and we assume a Gaussian distribution for the
  value of $\epsilon'_{\scalebox{0.5}{0}j}$ with a mean value of zero and \textcolor{black}{a} standard
  deviation of $\delta \epsilon '_{\scalebox{0.5}{0}}$. We explain \textcolor{black}{the
  above in detail} in the
  supplementary materials.
  


{\bf{Comparison of experiments and simulations}}
\newline \textcolor{black}{ {\bf{Dispersive frequency shift}.}}
For sample A,
we can reproduce the experimental spectroscopic results \textcolor{black}{with} our model as
shown in the Fig. \ref{samplea}a. 
From the \textcolor{black}{modeling},
the mean value
(standard deviation) of the tunneling energy of
the flux qubits is estimated as $\bar{\Delta }/2\pi =9.74$ GHz ($\sigma
_{\Delta }/2\pi=1.7 $ GHz). 
The negative resonator-frequency shift of
$\delta \omega_{\text{res}}/2\pi \simeq $250 MHz can be understood
as the dispersive energy shift \cite{blais2007quantum,wallraff2004strong}.
Since this
experiment is implemented in a dilution refrigerator with a temperature 
of 50 mK, the flux qubit is prepared in the ground state as long as
the qubit energy is much larger than the thermal energy of $k_BT/2\pi
\hbar \simeq 1$ GHz.
When most of the flux qubit energy is well above the resonator frequency,
each qubit induces
a negative resonator frequency shift of $-\frac{|g_j|^2}{\omega _j -\omega
_{\text{r}}}$.
Due to a collective effect, we can achieve a large dispersive shift
\textcolor{black}{of} $\delta \omega_{\text{res}}/2\pi\simeq $250 MHz
\textcolor{black}{for sample A}.
Although the individual coupling $\overline{g}'/2\pi (=14.3\rm{ MHz})$ is small, the collective effect
enhances the coupling strength $\sqrt{N}$ times \cite{imamouglu2009cavity,wesenberg2009quantum}.
Also, the resonator width tends to \textcolor{black}{becomes} larger as
the flux qubits approaches the degeneracy point for $\epsilon \simeq
0$  (see Fig. \ref{samplea}a.)
This is reasonable because the detuning between the flux qubit and the
 resonator becomes smaller as the operating point of the flux qubits
 approaches to the
 degeneracy, and this should induce additional decay in the resonator.

Importantly, these experimental results provide an order estimation of
the number of flux qubits coupled with the resonator.
The bare coupling strength between a single flux qubit and the resonator is described as $g'_j=M_{\text{qr}}I_j
\sqrt{\frac{\omega _{\text{r}}}{2\hbar L}}$ where $L$ denotes
the inductance of the resonator and $M_{\text{qr}}$ denotes a
mutual inductance between the flux qubit and the resonator. We can estimate these values as $M\simeq
10$ pH
and $L\simeq 100$ nH from numerical simulations, and so we obtain
$g'_j/2\pi \simeq 16 $ MHz for $\omega _{\text{r}}/2\pi =5.5$ GHz and
$I_j=250$ nA.
On the other hand, by reproducing the spectroscopic measurements in Fig. \ref{samplea}, we estimate the
average bare coupling strength as
$\bar{g}'/2\pi \simeq 14.3$ MHz where we
assume \textcolor{black}{$N=4300$}. This small
discrepancy in the estimated coupling strength of $g'_j$ might indicate that,
although we \textcolor{black}{intended} to fabricate \textcolor{black}{4300} flux qubits, some of them would not work as 
qubits \textcolor{black}{because of} imperfect fabrication. This could mean that the actual number of flux
qubits \textcolor{black}{contributing to} the collective enhancement might be smaller than \textcolor{black}{$4300$}.
However, from these estimations, we can at least conclude that thousands of
flux qubits should be involved in the collective coupling with the resonator, because
otherwise experimental results such as the large dispersive shift of $\delta
\omega_{\text{res}}/2\pi \simeq $250
MHz cannot be explained by the parameters $N<1000$ and $g'_j/2\pi \simeq 16 $
MHz.

\textcolor{black}{{\bf {Thermal effects on the energy shift.}}}
For sample B,
we can reproduce the experimental spectroscopic results with our theoretical model as
shown in Fig. \ref{sampleb}b.
From these theoretical calculations, the mean value
(standard deviation) of the tunneling energy of
the flux qubits is estimated \textcolor{black}{to be} $\bar{\Delta }/2\pi =1.17$ GHz ($\sigma
_{\Delta }/2\pi=1.1$ GHz). 
Both negative and positive frequency
shifts of the resonator can occur
when the
tunneling energy of the flux qubit is smaller than the resonator
frequency. In this case, the flux-qubit energy can cross the resonator frequency by
applying a magnetic field.
Unfortunately, due to a large inhomogeneous broadening of a few GHz, we
cannot observe the vacuum Rabi splitting of \textcolor{black}{the} qubit-resonator
anticrossing.
However, the positive and negative energy shifts indicate the existence
of coupling between the flux-qubit ensemble and the resonator.
Interestingly, the dispersive shift  around the degeneracy point for sample B is $\delta
\omega_{\text{res}}/2\pi \simeq $16 MHz, which is much smaller than the shift observed in sample A.
This is due to the small tunneling energy ($\bar{\Delta }/2\pi  \sim 1$ GHz) of
the flux qubit where the thermal energy depolarizes the flux qubit, which weakens
the dispersive shift.

We can reproduce the temperature dependence of the
spectroscopic measurements \textcolor{black}{with} our model as
shown in Fig. \ref{samplea}c.
 The temperature dependence becomes clearer
as the flux qubit approaches
the degeneracy point. Since the energy of the flux qubit
\textcolor{black}{reaches its} minimum at
the degeneracy point, the state of the flux qubit strongly
depends on the environmental temperature \textcolor{black}{because of} the small tunneling
energy, and this induces a temperature
dependent dispersive shift. 
On the other hand, far from the degeneracy point,
the resonator frequency shift becomes almost independent \textcolor{black}{of} the
temperature, because the flux qubit is not significantly affected by the thermal
effect \textcolor{black}{caused by} the large flux qubit energy.


\textcolor{black}{{\bf {Discrete nature of the qubits.}}}
Interestingly, on top of the dispersive shift, we observe numerous
small energy shifts in the spectroscopy in the
experiments and simulations (see 
Fig. \ref{samplea}c).
\textcolor{black}{This is because a large finite number of flux qubits
are coupled with the resonator.}
If a single qubit was coupled with the resonator, only
the dispersive shift and/or the vacuum Rabi splitting
should be observed in the spectroscopy \textcolor{black}{because of} the change
of the eigenenergy of the resonator coupled
with a qubit
\cite{blais2007quantum,wallraff2004strong}.
On the other hand, in the limit of a large
number of qubits coupled with the resonator, we
 should also observe the dispersive frequency
shift and/or vacuum Rabi splitting in the spectroscopy, because we can
consider the qubit ensemble \textcolor{black}{to be} a single harmonic
oscillator as a consequence of the continuum limit \cite{diniz2011strongly}.
In the experiments \textcolor{black}{described} here, since the resonator is coupled
with thousands of
qubits, which is \textcolor{black}{large} but finite \textcolor{black}{number}, we observe numerous
additional energy shifts in the spectroscopy due to the discrete nature
of each flux qubit.

{\bf{Discussion}}
\newline
In conclusion, we have reported experiments \textcolor{black}{that} show collective
coupling between a superconducting resonator and an ensemble of
superconducting flux qubits.
We have observed \textcolor{black}{the} large dispersive frequency
shift of the resonator, and this demonstrates
a collective behavior of
the superconducting flux qubits.
A quantitative analysis indicates that thousands of superconducting
flux qubits contribute to the collective coupling.
These results represent the largest number of coupled superconducting
qubits realized so far.
Our system \textcolor{black}{has} many potential applications \textcolor{black}{including} quantum metamaterials,
quantum metrology, and a quantum simulator.

{\bf{Methods}}
\newline
{\bf{ Experimental setup.}}
\begin{figure}[h] 
\begin{center}
\includegraphics[width=0.9\linewidth]{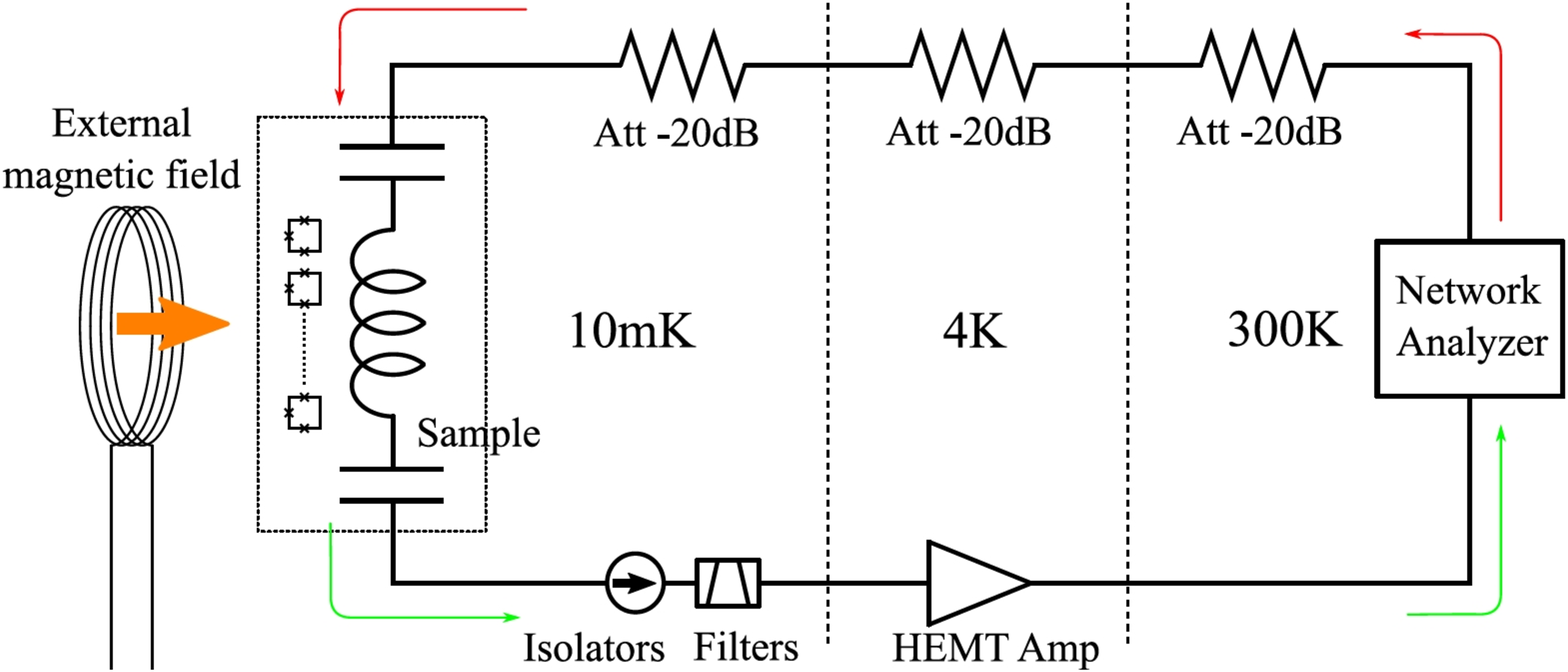} 
\caption{
 Schematic of our
 experiment setup. Here, we measure the microwave transmission properties of
 the resonator
 with a network analyzer. The microwave resonator is inductively coupled
 with 4500 flux qubits.
 }
 \label{setup}
\end{center}
\end{figure}

 We fabricate a microwave resonator and 4300 flux qubits on a Si wafer
 by using electron beam lithography and conventional angled evaporation
 technique.
 Here, the Si wafer has SiO$_2$
 thin film.
 Our flux qubit has three Al-Al$_2$O$_3$-Al Josephson junctions.

Our experimental setup is described in the Fig. \ref{setup}.
We measure the microwave transmission properties of the resonator
system around 6 GHz by a network analyzer. 
To amplify the output microwave signal, we prepare HEMT amplifier on 4K stage.
We insert attenuators on each temperature stage. Also, we insert band pass filters and isolators to avoid thermal noise coming from room temperature and amplifier.
We can apply magnetic fields perpendicular to qubits to change the operating points of the superconducting flux qubits.

We can change the temperature from
10 mK to 230 mK.
To measure the temperature,
we can use a RuO$_2$ thermometer at a mixing chamber, and we stabilize it
by using a PID feedback controlled heater.


{\bf{Theoretical model.}}
We describe the theoretical model \textcolor{black}{that we used for the
results described} in the main text.

Firstly, we derive a formula \textcolor{black}{for} the transmitted photon intensity when
the effect of the thermal energy is negligible.
If the excited state of the flux qubit is not significantly populated
with weak microwave driving,
we can approximate the flux qubit as a harmonic oscillator \cite{houdre1996vacuum}.
Here, only
the lowest two levels of the harmonic oscillator are occupied, and the
effect of the other
levels is negligible due to the weak driving power. 
With the replacement of $\hat{\sigma }^+_j\rightarrow \hat{b}^{\dagger }_j$,  we can represent Heisenberg equations based on the Hamiltonian in Eq. \ref{hamiltonian}
\begin{eqnarray}
&&i\hbar \frac{d\hat{a}}{dt}=\hbar \lambda +\hbar
 (\omega_{\text{r}}-\omega-i\gamma_{\text{r}})\hat{a}+\sum\limits_{j} \hbar
 g_{j}\hat{b}_{j},\nonumber \\
&&i\hbar\frac{d\hat{b}_{j}}{dt}=\hbar(\omega_{j}-\omega-i\gamma_{\text{qubit}})\hat{b}_j+\hbar g_{j}\hat{a}\nonumber
\end{eqnarray}
We use a steady condition such as $\frac{d\hat{a}}{dt}=0$ and $\frac{d\hat{b}_{j}}{dt}=0$ and we obtain
\begin{eqnarray}
 \hat{a}&&=\label{lowt}\frac{\lambda}{\omega-\omega_{\text{r}}+i\gamma_{\text{r}}-\sum\limits_{j} \frac{g_{j}^2}{{\omega-\omega_{j}+i\gamma_{\text{qubit}}}}}.
\end{eqnarray}
In a weak coupling regime, we can assume
$\vert\omega-\omega_{\text{r}}\vert\ll\vert\omega_{\text{r}}-\omega_{j}+i\gamma_{j}\vert$
for $j=1,2,\cdots ,N$, and so we obtain
\begin{eqnarray}
 \hat{a}&&\simeq \frac{\lambda }{(
  \omega-(\omega_{\text{r}}+\delta\omega_{\text{r}}))+i(\gamma_{\text{r}}+\delta
  \gamma_{\text{r}})}\nonumber \\
 \delta\omega_{\text{r}}&=&-\sum\limits_{j}
  g_{j}^2\frac{\omega_{j} -\omega_{\text{r}}}{(\omega_{j}-\omega_{\text{r}})^{2}+\gamma_{\text{qubit}}^{2}}\nonumber \\
  \delta\gamma_{qubit}&=&\sum\limits_{j}
   g_{j}^2\frac{\gamma_{\text{qubit}}}{(\omega_{j}-\omega_{\text{r}})^{2}+\gamma_{\text{qubit}}^{2}}\nonumber \\
\end{eqnarray}

Secondly, we derive
a formula \textcolor{black}{for} the resonator
 frequency shift induced by
 the detuned flux qubits whose energies 
 are comparable \textcolor{black}{to} the thermal energy.
Due to the energy difference between the microwave resonator and
 \textcolor{black}{the} flux qubits,
 we can use a dispersive Hamiltonian to describe this system as
\begin{eqnarray}
&&H_{\text{disp}}\simeq \hbar \sum\limits_{j}( \frac{\omega_{j}-\omega }{2}+\frac{g_{j}^2}{2(\omega_{j}-\omega_{\text{r}})})\hat{\sigma}_{z,j}\nonumber \\
 &+& \hbar (\omega_{\text{r}}-\omega +\sum\limits_{j}
  g_{j}^2\frac{\hat{\sigma}_{z,j}}{\omega_{j}-\omega_{\text{r}}})\hat{a}^{\dagger}\hat{a}+\hbar
  \lambda (\hat{a}+\hat{a}^{\dagger })\ \ \ \ \ \ 
  \label{detune}
\end{eqnarray}
In this case, we have a dispersive shift for the energy of the
microwave resonator such as
\begin{eqnarray}
 \delta\omega_{\text{r}} &&=\sum\limits_{j} g_{j}^2\frac{\langle\sigma_{z,j}\rangle}{\omega_{j}-\omega_{\text{r}}}
\end{eqnarray}
where $\langle\hat{\sigma}_{z,j}\rangle$ denotes the expectation value
of $\hat{\sigma}_{z,j}$.
Since the flux qubit is in a thermal equilibrium, we obtain
\begin{eqnarray}
 \delta\omega_{\text{r}} &&=-\sum\limits_{j} g_{j}^2\frac{\text{tanh}(\frac{\hbar \omega_{j}}{ 2 k_{B} T_{\text{E}}}) 
  }{\omega_{j}-\omega_{\text{r}}}\label{secondregime}
\end{eqnarray}
Also, since the energy relaxation rate of a qubit with an energy of
$\omega _j$ is proportional to $(1+\overline{N})$
where $\overline{N} =(e^{\frac{\hbar \omega
_j}{k_BT_{\text{E}}}}-1)^{-1}$ denotes the Bose-Einstein occupation
number of the environmental bosonic modes \cite{GZ01b}, we assume
\begin{eqnarray}
  \gamma
_{j}&=& 
 (1+(e^{\frac{\hbar \omega
_j}{k_BT_{\text{E}}}}-1)^{-1} )\gamma _{\text{qubit}}
\end{eqnarray}
and so we can take the thermal effect into account in the decay rate of
the flux qubit.

Therefore, from the equations described above, we derive the following phenomenological function 
\begin{eqnarray}
 \langle \hat{a}^{\dagger }\hat{a}\rangle\simeq\frac{\lambda^2}{(\omega-(\omega_{\text{r}}+\delta\omega_{\text{r}}))^2+(\gamma_{\text{r}}+\delta \gamma_{\text{r}})^2} \ \ 
\end{eqnarray}
where we have
\begin{eqnarray}
 \delta\omega_{\text{r}} &&=-\sum\limits_{j} g_{j}^2\frac{\text{tanh}(\frac{\hbar \omega_{j}}{ 2 k_{B} T_{\text{E}}})(\omega_{j}-\omega_{\text{r}})}{(\omega_{j}-\omega_{\text{r}})^{2}+\gamma_{j}^{2}}\ \ \ 
\end{eqnarray}
\begin{eqnarray}
 \delta\gamma_{qubit}=\sum\limits_{j} g_{j}^2\frac{\gamma_{j}}{(\omega_{j}-\omega_{\text{r}})^{2}+\gamma_{j}^{2}}.
\end{eqnarray}
This function will coincide with 
Eq. (\ref{lowt}) or (\ref{secondregime}) in the limit of a low temperature or a
large detuning.
Since the number of photons inside the resonator corresponds to the
intensity of the transmitted photons \cite{diniz2011strongly}, we can
use this function to reproduce the experimental results.
It is worth mentioning that we observed experimentally the applied magnetic flux
dependence of the microwave driving amplitude $\lambda =\lambda (
\Phi  _{\text{ex}})$. Although this cannot be
explained by our theoretical model, we obtain the shape of $\lambda
(\Phi  _{\text{ex}} )$ from the experiment, and substitute this into our theoretical
model to reproduce the spectroscopic measurements (See the 
supplementary materials for details).

{\bf {Acknowledgments}}
\newline
We thank S. Endo, I. Natsuko, and N. Lambert for their useful comments.
This work was supported by JSPS KAKENHI Grant
15K17732, JSPS KAKENHI Grant No.25220601, the Commissioned Research No. 158 of
NICT, and MEXT KAKENHI Grant Number 15H05870.

{\bf {Author contributions}}
\newline
All authors contributed extensively to the work presented in this paper.
K. K fabricated the device and carried out measurements.
Y.M., C.D., and W.J.M provided theoretical support and analysis.
Y.M. wrote the manuscript, with feedback from all authors.
H.Y., W.J.M. and S.S. supervised the project.

\section{Supplementary Figures}
\begin{figure}[!h] 
\begin{center}
\includegraphics[scale=0.53]{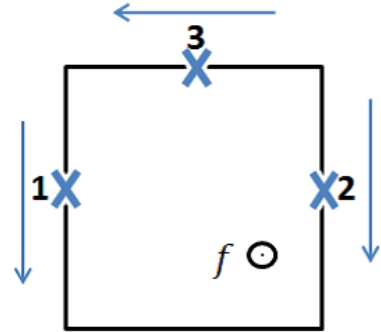} 
\caption{
 Schematic of a superconducting flux qubit
 with three Josephson junctions.
   The area of one junction is designed to be $\alpha $ times smaller than
those of the other two junctions.
 }
 \label{fluxqubit}
\end{center}
\end{figure}
\begin{figure*} 
\begin{center}
\includegraphics[scale=0.2]{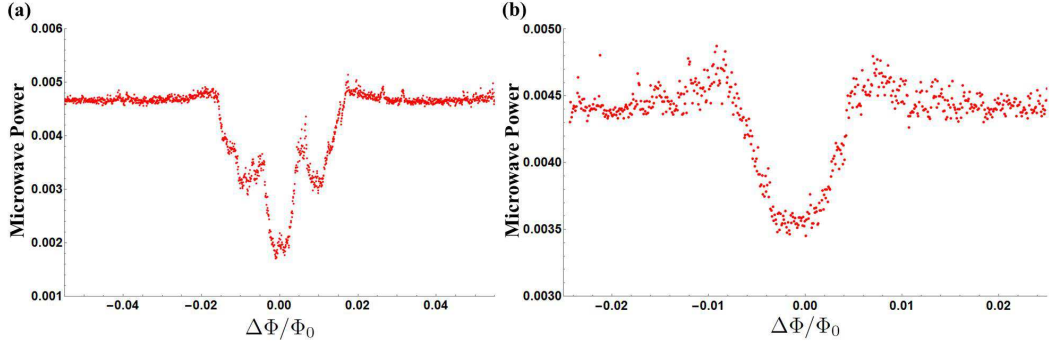} 
\caption{
 Applied magnetic flux dependence of the microwave amplitude in our experiment.
 For a fixed amount of applied magnetic flux, we fit the
 experimental spectroscopic results (shown in Fig. 2 in the main text) by \textcolor{black}{using}
 $f(\omega )=\frac{\Lambda ^2 }{(\omega -\omega _{\text{av}}+\gamma^2)}$
 and obtain the fitting parameters of $\Lambda $, $\omega _{\text{av}}$,
 and $\gamma $. We plot $\Lambda$ here for sample A (left) and
 sample B (right). These \textcolor{black}{plots} show that the microwave driving amplitude
 \textcolor{black}{is dependent} on the applied magnetic flux.
 }
 \label{power}
\end{center}
\end{figure*}

\newpage

\section{Supplementary Note 1: Inhomogeneity of Josephson Junctions}

We describe how we take \textcolor{black}{the} inhomogeneity of the Josephson Junctions into
account in our model.
We consider a superconducting circuit with three Josephson junctions as
shown in the Fig. \ref{fluxqubit}
The Lagrangian of this system \cite{orlando1999superconducting} can be written as
\begin{eqnarray}
 L&=&T-U \\
 U&=&\sum\limits_{j=1}^{3} (
\frac{\Phi_{0}}{2\pi}I_{\text{C}}^{j}(1-\text{cos}(\phi_{j})) \\
 T&=&\sum\limits_{j=1}^{3} \frac{1}{2}C_{j}(\frac{\Phi_{0}}{2\pi})^{2}\dot{\phi_{j}}^{2}
\end{eqnarray}
where $U$ denotes the total potential energy, $T$ denotes the total
kinetic energy, $\phi_{j}$ $(j=1,2,3)$ denotes
the phase difference between the junctions, $C_{j}$ denotes the
capacitance of the Josephson junction, $I_{\text{C}}^{j}$ denotes
 the critical current,  $\Phi_{\text{ext}}$ denotes
 the flux due to the external magnetic field, and
 $\Phi_{0}=\frac{\hbar}{2e}$ denotes the magnetic flux quantum.
 $E^{(j)}_J=\frac{\Phi_{0}}{2\pi}I_{\text{C}}^{j}$ denotes the
 characteristic scale of the Josephson energy while
 $E_c=\frac{e^2}{2C_j}$ sets the characteristic electric energy.
We have a condition $\phi_{1}-\phi_{2}+\phi_{3}=2\pi f$ with
$f=\frac{\Phi{\text{ext}}}{\Phi_{0}}$.
It is worth mentioning that $C_j$ and $I_{\text{C}}^{j}$ are
proportional to the size of the junction.
  \textcolor{black}{The} area for one junction is $\alpha $ times smaller than
those for the other two junctions if the device is fabricated as
designed. 
However, Josephson junctions are very sensitive \textcolor{black}{to} experimental
conditions, and the size of the junctions becomes inhomogeneous.
  To consider \textcolor{black}{this} inhomogeneity, we assume a Gaussian distribution for normalized areas of the smaller
  junction (two larger
  junctions)
  with a mean value of $\bar{\alpha }$
  ($\bar{\beta }_k$ for $k=1,2$) and a standard
  deviation of $\sigma _{\text{S}}$ ($\sigma ^{(k)}_{\text{L}}$ for
  $k=1,2$).
  By solving this, we can calculate the energies of the
  ground state and excited state for a given external magnetic flux,
  which provides us with the values of the tunneling energy
$\Delta _j$ and the persistent current $I_j$ of the flux qubit \cite{orlando1999superconducting}.



 \section{Supplementary Note 3: Applied magnetic flux dependence of microwave
 driving amplitude}

 In the spectroscopic measurements shown in Fig. 2 in the main text,
 we experimentally observe a Lorentz distribution for a fixed amount of
 applied magnetic flux. We fit the spectrum by a function
 of $f(\omega )=\frac{\Lambda ^2 }{(\omega -\omega _{\text{av}}+\gamma
 ^2)}$, and obtain the fitting parameters of $\omega _{\text{av}}$,
 $\gamma $, and $\Lambda $. Our theoretical model described in the
 main text predicts that the microwave amplitude \textcolor{black}{will be} independent \textcolor{black}{of} the
 applied magnetic flux. However, the
 $\Lambda $ value obtained from the fitting has some dependence on
 the applied magnetic flux as shown in Fig. (\ref{power}). 
 Although this cannot be
explained by our theoretical model, we
substitute this into our theoretical
model ($\lambda =\Lambda $) to reproduce the spectroscopic measurements.

A possible reason
\textcolor{black}{is that the
properties of the flux qubit change due to the applied magnetic field, and this may also change the impedance of
the resonator. Such variations in the impedance induce additional reflection of
the input photon, and this could explain the dependence of the driving
strength on the applied magnetic fields.}


\end{document}